\documentclass[aps,twocolumn,amsmath,amssymb]{revtex4}
\usepackage{graphicx, graphics,epsfig}

\newcommand{\be}{\begin{equation}}
\newcommand{\ee}{\end{equation}}

\begin{document}
\setcounter{page}{145}

%%%%%%%%%%%%%%%%%%%%%%%%%%%%%%%%%
%%%%%%%%%%%%%%%%%%%%%%%%%%%%%%%%%
{\widetext
\begin{center}
{\small CONTRIBUTIONS \hskip1cm REVISTA MEXICANA DE F\'ISICA {\bf
49 SUPLEMENTO 2}, 145--147 \hskip1cm OCTUBRE 2003}
\end{center}
}
%%%%%%%%%%%%%%%%%%%%%%%%%%%%%%%%%
%%%%%%%%%%%%%%%%%%%%%%%%%%%%%%%%%

\title{On susy periodicity defects of the Lam\'e Potentials}

\author{Oscar Rosas-Ortiz}

\affiliation{\ Departamento de F\'{\i}sica, CINVESTAV-IPN, A.P. 14-740,
07000 M\'exico D.F., Mexico\\
{\rm Recibido el 28 de marzo de 2002; aceptado el 19 de junio de 2002}
}

\begin{abstract}
The supersymmetric Quantum Mechanics approach is applied to embed bound
states in the energy gaps of periodic potentials. The mechanism to
generate periodicity defects in the first Lam\'e potential is analyzed.
The related bound states are explicitly derived.
\end{abstract}

\maketitle

%%%%%%%%%%%%%%%%%%%%%%%%%%%%%%%%%%
\vskip0.5cm
%\begin{multicols}{2}

Let $H=-\frac{d^2}{dx^2} + V(x)$ be a Schr\"odinger operator with $V(x)$ a
real-valued and locally bounded function on ${\mathbb R}$. The spectrum
$\sigma(H)$ of $H$, as well as its resolvent set $\rho(H)$, will be then
complementary subsets of ${\mathbb R}$. The involved eigenvalue equation
$H\,
\psi(x) = \epsilon \psi(x)$, admits solutions for arbitrary $\epsilon \in
{\mathbb R}$. The $\epsilon$-values such that $\epsilon =E \in
\sigma_d(H)$
lead to square integrable wave-functions $\psi_E(x)$ in $L^2({\mathbb 
R})$,
while for $\epsilon \in \rho(H)$ one gets eigenfunctions $\psi(x)$
deprived of physical meaning. It is well known that the class of
Schr\"odinger's operators $H$ admitting analytic expressions for the
corresponding eigenfunctions is narrow. The supersymmetric (susy) Quantum
Mechanics \cite{Coo00} intends to expand the class. In particular, the
conventional Infeld-Hull factorization method has been extended to employ
not only the physically interpretable $\psi_E$ but also the nonphysical
eigenfunctions $\psi$ \cite{Mie84}. The main idea of this generalized
factorization is to implement the Darboux-Crum transformation as the
mechanism to construct multi-parametric families $\{ V(\lambda_1,
\lambda_2, ..., \lambda_n,x)\}$ of (almost) isospectral exactly solvable
potentials \cite{everybody,Fer97}. The members of each one of these
families share common analytical properties, such as the number and type
of (possible) singularities as well as the asymptotic behaviour. Hence, it
is intended that the procedure does not add new singularities nor modifies
the asymptotic behaviour of the initial potentials. This mechanism is
ensured by taking eigenfunctions $\psi$ with $\epsilon \in \rho(H)$ as the
key transformation functions \cite{Fer97}.

The present contribution deals with the generalized factorization method
applied to periodic potentials. As we shall see, the very interesting
phenomenon of periodicity defects in crystal lattices can be produced as a
consequence of the susy mechanism. We support our analysis mainly on the
results reported recently in \cite{Fer02}.

Let us consider a Schr\"odinger operator $H_0$ for which the solutions
$u(x,\epsilon)$ to the eigenvalue equation 
\be 
H_0 u(x,\epsilon) =\epsilon u(x,\epsilon) 
\label{eigen1} 
\ee 
are known for any $\epsilon \in {\mathbb R}$. Notice that equation
(\ref{eigen1}) does not require the square integrability of
$u(x,\epsilon)$. An important example is the Hamiltonian with a
real-valued, continuous and periodic potential
\be
V_0(x+ \tau) = V_0(x), \qquad x \in {\mathbb R}. 
\label{periodic} 
\ee 
In this case, the translation operator $T u(x,\epsilon) = u(x + \tau,
\epsilon)$, maps the two-dimensional subspace ${\cal H}_{\epsilon}$ of
solutions of the eigenvalue equation (\ref{eigen1})  into itself. As a
consequence, there is a basis of eigenfunctions $\{ {\rm v}_1(x), {\rm
v}_2(x) \}$ in ${\cal H}_{\epsilon}$ for which $T$ admits a $2 \times 2$
matrix representation $[T]$ (the {\it monodromy matrix\/}). The eigenvalue
equation for $[T]$, evaluated at $x=T$, determines two eigenvalues
$t_{\pm}$ for a given energy $\epsilon$, i.e., $t_{\pm} =
t_{\pm}(\epsilon)$. The corresponding eigenfunctions $v_{\pm}(x,
\epsilon)$ exist for any $\epsilon \in {\mathbb R}$ and are called {\it
Bloch
functions\/}. The structure of $v_{\pm}(x, \epsilon)$ depends essentially
on the values of $\epsilon$, namely, if $\epsilon$ is such that $t_{\pm}
\in S^1$, then $\epsilon \in \sigma(H_0)$ and the related Bloch functions
become complex and bounded. The spectrum of $H_0$ is then continuous,
composed by a series of spectral bands in ${\mathbb R}$. The edges of that
bands are given by the $\epsilon$-values such that $t_{\pm}=\pm 1$, and
the Bloch functions become $\tau$-periodic or $\tau$-antiperiodic.
Finally, if $\epsilon$ is such that $t_{\pm} \in {\mathbb R}/\{\pm1\}$,
then
$\epsilon \in \rho(H_0)$ and the related Bloch functions are real and
diverge at either $-\infty$ or $+\infty$. Due to this mechanism,
$\rho(H_0)$ splits into a (finite or infinite) sequence of forbidden gaps.  
In summary, $\sigma(H_0)  = \{ \epsilon \, ; \, t_{\pm} \in S^1\}$, while
$\rho(H_0) = {\mathbb R}/\sigma(H_0)$.

As regards to the supersymmetric treatment of periodic Hamiltonians, one
can consider two different options: (i) The Darboux transformations are
such that, when applied to systems described by (\ref{periodic}), the new
potentials are also $\tau$-periodic. (ii) The procedure does not respect
the global periodicity of the initial potential. In the former case, Dunne
and Feinberg \cite{Dun98} have shown that, sometimes, the nonsingular
first order transformed potential becomes simply a half-period displaced
copy of the initial one. Such potentials were called {\it
self-isospectral\/}. A further study (using the band edge eigenfunctions)
produced the 2-susy self-isospectral potentials \cite{Fer00}. The above
results have been recently generalized to embrace arbitrary
$\delta$-displaced copies of the initial periodic potential \cite{Fer02}.
Such a phenomenon has been named the {\it translational Darboux
invariance\/} (or simply Darboux invariance), and it reproduces the
self-isospectral case for $\delta = \tau/2$. The Darboux invariance is
achieved by using Bloch solutions as transformation functions for
$\epsilon$ either at the band edges or in the resolvent set of the
periodic Hamiltonian $H_0$. Concerning the option (ii), we first remark
that arbitrary linear combinations of $v_{\pm}(x, \epsilon)$ for $\epsilon
\in \rho(H_0)$, when used as transformation functions $u(x, \epsilon)$ can
produce regular non-periodic modifications in the periodic potential
(\ref{periodic}). As we are going to show, moreover, the functions $u(x,
\epsilon)$ allow to embed discrete energy levels in the gaps of $H_0$ and
to determine the corresponding square integrable eigenfunctions
\cite{Fer02}.

The {\it generalized\/} first order factorization method \cite{Mie84}
allows to obtain the solutions $\phi_{\epsilon'}(x,\epsilon)$ of a
transformed eigenvalue equation $H_1(\epsilon) \phi_{\epsilon'} (x,
\epsilon) = \epsilon' \phi_{\epsilon'} (x,\epsilon)$; \, $H_1(\epsilon):=
-\frac{d^2}{dx^2} + V_1(x,\epsilon)$, where the new potential
$V_1(x,\epsilon)$ is related with the initial one $V_0(x)$ by means of a
Darboux transformation
\be
V_1(x,\epsilon) = V_0(x) + 2 \beta'(x,\epsilon).
\label{darboux1}
\ee
The function $\beta$ here is nothing but the general solution of the
Riccati equation
\be
-\beta'(x, \epsilon) + \beta^2(x, \epsilon) = V_0(x) - \epsilon.
\label{riccati1}
\ee
At this stage, it is interesting to notice that the transformation
$\beta(x,\epsilon) \equiv - \frac{d}{dx} \ln u(x, \epsilon)$ allows to
simplify the expressions for the potential (\ref{darboux1}) and the
involved eigenfunctions $\phi_{\epsilon'}(x,\epsilon)$ leading to:
\be
\begin{array}{cl}
V_1(x, \epsilon) = V_0(x) -2 \frac{d^2}{dx^2} \ln u(x,\epsilon), &
\nonumber \\[2ex]
\phi_{\epsilon'}(x,\epsilon) \propto \frac{W(u, \psi_{\epsilon'})}{u(x,
\epsilon)} &
\end{array}
\label{susy}
\ee
where, by simplicity, we have written $u \equiv u(x, \epsilon)$ and
$\psi_{\epsilon'} \equiv \psi_{\epsilon'} (x)$. The symbol $W(u,
\psi_{\epsilon'})$ represents the Wronskian of $u$ and $\psi_{\epsilon'}$.

The analytical behaviour of $\phi_{\epsilon'}(x,\epsilon)$ strongly
depends on the values of $\epsilon$ and $\epsilon'$. If $\epsilon' = E \in
\sigma(H_0)$ and $E_0 > \epsilon \in \rho(H_0)$, with $E_0 \in
\sigma(H_0)$ denoting the lowest band edge, the eigenfunction
$\phi_E(x,\epsilon)$ can be bounded. A particular detail of the method is
that the function $\phi_{\epsilon}(x, \epsilon) \propto 1/u(x, \epsilon)$
satisfies identically $H_1(\epsilon)  \phi_{\epsilon}(x, \epsilon) =
\epsilon \phi_{\epsilon}(x, \epsilon)$ and becomes orthogonal to all the
$\phi_E(x, \epsilon)$. Moreover, it can be chosen in $L^2({\mathbb R})$.
Thereby $\sigma(H_1(\epsilon)) = \sigma(H_0) \cup \{ \epsilon \}$. In
other words, the Hamiltonian $H_1(\epsilon)$ can be constructed in order
to have a bound state embedded in the lowest spectral gap of $H_0$. Hence,
$H_1(\epsilon)$ will be almost isospectral with $H_0$ for a properly
selected $\epsilon$. This situation allows to consider them as susy
partner Hamiltonians. Concerning the analytical properties of $V_1(x,
\epsilon)$, it is very common to find that those conditions on $\epsilon$,
making $\phi_E (x, \epsilon)$ a square integrable function, lead to
potentials (\ref{darboux1}) with exactly the same number and nature of
divergences as the initial one (see for example \cite{Fer97}). In
particular, as for $\epsilon < E_0$ the function $u(x,\epsilon)$ can be
chosen nodeless, the potential $V_1(x,\epsilon)$ in (\ref{susy}) is just a
regular non-periodic modification of $V_0(x)$. The Figure~1 shows one of
these first order regular deformations for the simplest Lam\'e potential
\be
V_0(x) = 2 m \, {\rm sn}^2 (x \vert m).
\label{lame}
\ee
Let us remark that the functions $u(x, -0.1)$, producing the potential
$V_1(x, -0.1)$ in Figure~1, preserve the smoothness of (\ref{lame}) as
well as its asymptotic behaviour. Moreover, the deformation of $V_1$ in
the neighbourhood of $x=0$ indicates the presence of the bound state
energy level at $\epsilon =-0.1$ in the first forbidden gap $(-\infty,
E_0)$. For the present case, the related wave-function $1/u(x, -0.1)$ is a
smooth function of $x$, sharply peaked at $x=0$.

\medskip
%%%%%%%%%%%%%%%%%%%%%%%%%
\begin{figure}[htbp]
\centering \epsfig{file=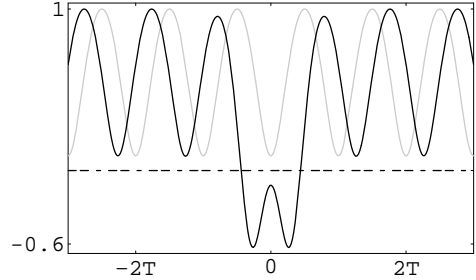, width=7cm}
%\begin{minipage}{8.5cm} 
%\end{minipage}
%\begin{figure}
%\graphicspath{/home/orosas/PROCEED/proc12}
%\DeclareGraphicsExtensions{eps,ps}
%\includegraphics[width=7cm,height=7cm]{orof1}
\caption{The Lam\'e potential (\ref{lame})  with
$m=0.5$ (gray) and its first order regular deformation (black) with a
bound state energy level at $\epsilon=-0.1< E_0=0.5$ (dashed).}
\end{figure}
%%%%%%%%%%%%%%%%%%%%%%%%% 
\medskip

\noindent
The procedure can be now iterated by considering $\epsilon = \epsilon_a$
as given and $V_1(x,\epsilon_a)$ as the known exactly solvable potential.
Thus, one has to look for a new potential $V_2(x, \epsilon_a, \epsilon)$
satisfying the second order relationship
\be
V_2(x,\epsilon_a, \epsilon) = V_1(x, \epsilon_a) +
2 \beta'(x,\epsilon_a, \epsilon).
\label{darboux2}
\ee
Therefore, the new function $\beta(x,\epsilon_a, \epsilon)$ must fulfill
the Riccati equation $-\beta'(x, \epsilon_a, \epsilon) + \beta^2(x,
\epsilon_a, \epsilon) = V_1(x, \epsilon_a) - \epsilon$, where $\epsilon$
is arbitrary. The solutions of this equation are easily determined by
knowing those of (\ref{riccati1}) for at least two different factorization
constants $\epsilon$, $\epsilon_a$ \cite{Fer97}. Then, one can use the
finite-difference B\"acklund algorithm \cite{Fer98}. It is remarkable that
$\beta(x, \epsilon_a, \epsilon)$ appears to be symmetric under the
interchange of $\epsilon_a$ by $\epsilon$ and vice versa. This means that
one can achieve the potential (\ref{darboux2}) by finding first $V_1(x,
\epsilon_a)$ and then $V_2(x, \epsilon_a, \epsilon)$ or, first $V_1(x,
\epsilon)$ and then $V_2(x, \epsilon, \epsilon_a)$ \cite{Fer97}.  More
details concerning further iterations and symmetry properties of the
Darboux transformations can be found in \cite{Nie01} and references quoted
therein. Observe that potential (\ref{darboux2}) and the resulting
eigenfunctions $\phi_{\epsilon'}(x,\epsilon_a, \epsilon)$ can be rewritten
as:
\be
\begin{array}{cc}
V_2(x, \epsilon_a, \epsilon) = V_0(x) -2 \frac{d^2}{dx^2} \ln W(u_a,u); &  
\nonumber \\[2ex]
\phi_{\epsilon'}(x,\epsilon_a, \epsilon) \propto \frac{W(u_a, u,
\psi_{\epsilon'})}{W(u_a,u)} &
\end{array}
\label{sususy}
\ee
where, for the sake of simplicity, we have written $u_a \equiv
u(x,\epsilon_a)$. Let us consider the precedent 1-susy step as given with
$\epsilon_a < E_0 \in \sigma (H_0)$. The Hamiltonian $H_1(\epsilon_a)$ is
therefore regular and $\phi_{\epsilon'}(x,\epsilon_a, \epsilon)$ becomes
an eigenfunction of $H_2(\epsilon_a, \epsilon)$ with eigenvalue
$\epsilon'$.  If $\epsilon' = E \in \sigma(H_0)$, $\epsilon \neq
\epsilon_a$ and $E_0 > \epsilon \in \rho(H_0)$, one easily finds that
$\phi_E(x,\epsilon_a, \epsilon)$ can be bounded. Yet, the function
$\phi_{\epsilon}(x,\epsilon_a, \epsilon) \propto 1/ \phi_{\epsilon} (x,
\epsilon_a)$ is also orthogonal to all the $\phi_E(x,\epsilon_a,
\epsilon)$ and can be chosen in $L^2({\mathbb R})$. Hence
$\sigma(H_2(\epsilon_a, \epsilon)) = \sigma(H_1(\epsilon_a)) \cup \{
\epsilon \}$ and the Hamiltonians $H_1(\epsilon_a)$, $H_2(\epsilon_a,
\epsilon)$ are almost isospectral. The iteration chain now establishes
that $H_2(\epsilon_a, \epsilon)$ is also a susy partner of $H_0$ of the
second order, {\it i.e.\/}, $H_2(\epsilon_a, \epsilon)$ and $H_0$ are now
``sususy partners''. In the present case, the iteration procedure leads to
nonsingular Hamiltonians at each step, hence, the sususy transformation is
said to be {\it reducible\/}.

\medskip
%%%%%%%%%%%%%%%%%%%%%%%%%
\begin{figure}[htbp]
\centering \epsfig{file=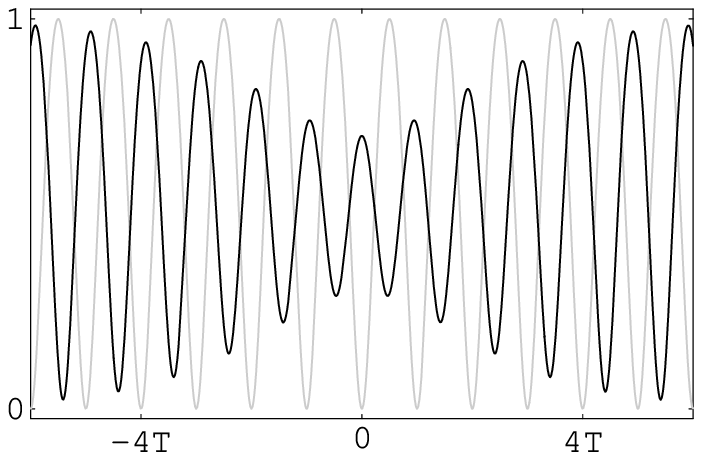, width=7cm}
\begin{minipage}{8.5cm}
{\bf Fig. 2.} {\footnotesize The initial Lam\'e potential (gray) and its
second order regular deformation (black) allowing two bound state energy
levels at $\epsilon_a= 1.1$ and $\epsilon=1.45$.  Observe the asymptotic
translational invariance of the potentials.}
\end{minipage}
\end{figure}
%%%%%%%%%%%%%%%%%%%%%%%%%
\medskip

Now, let us consider the case when $E_0 < \epsilon_a \in \rho(H_0)$. As
the function $u(x, \epsilon_a)$ in (\ref{susy}) has a finite number of
zeros in $(0, \tau)$, the potential $V_1(x, \epsilon_a)$, as well as the
functions $\phi_{\epsilon'}(x, \epsilon_a)$, become singular.  However,
there is still a chance to construct regular potentials $V_2(x,
\epsilon_a, \epsilon)$ and functions $\phi_{\epsilon'}(x, \epsilon_a,
\epsilon)$. It is enough to take $\epsilon_a, \epsilon \in \rho(H_0)$ such
that $E_j < \epsilon_a, \epsilon < E_{j'}$, with $E_j, E_{j'}$ two
subsequent band edges of $H_0$ (we have used the notation of
\cite{Fer00}). Hence, for an appropriate choice of the involved
parameters, the Hamiltonian $H_2(\epsilon_a, \epsilon)$ is regular
although the intermediary ones $H_1(\epsilon_a)$ and $H_1(\epsilon)$ are
singular. We then say that the second order transformation is {\it
irreducible\/}. Figure~2 shows the result of an irreducible second order
regular transformation of potential (\ref{lame}) for $m=0.5$. The values
of the $\epsilon$'s are in the gap $(E_1,E_1')=(1,1.5)$. On the other
hand, Figure~3 illustrates one of the related bound states for
$\epsilon=1.45$.

\medskip
%%%%%%%%%%%%%%%%%%%%%%%%%
\begin{figure}[htbp]
\centering \epsfig{file=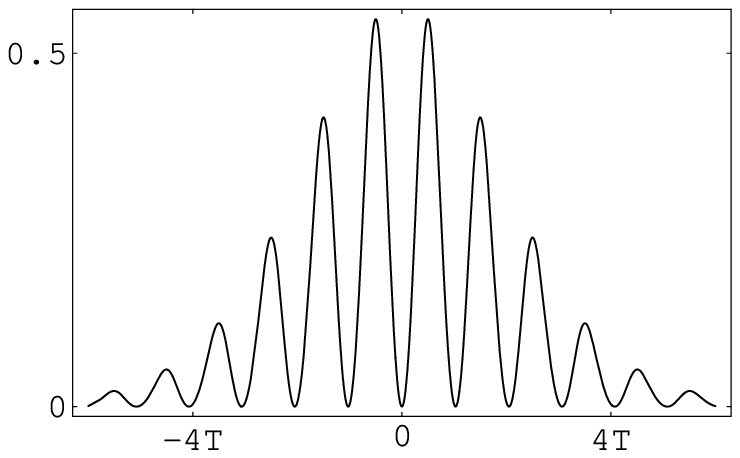, width=7cm}
\begin{minipage}{8.5cm}
{\bf Fig. 3.} {\footnotesize The probability density for one of the bound
states associated to the potential of Figure~2 at $\epsilon=1.45$}
\end{minipage}
\end{figure}
%%%%%%%%%%%%%%%%%%%%%%%%%
\medskip

\noindent Finally, let us remind that the presence of bound states in the
continuum is not a novelty in Quantum Mechanics. For example, Stillinger
and Herrick used a modification of the von Neumann-Wigner approach to
produce local potentials with bound eigenstates embedded in the continuum
of scattering states \cite{Sti75}. Similar results are obtained by means
of the susy Quantum Mechanics \cite{Pap93}. However, the situation when
the bound states are embedded in the energy gaps of a periodic potential
has been out of the attention for researches in the area.  Further details
on this subject can be found in \cite{Fer02}.

\bigskip
\noindent
The support of CONACyT project 32086E and of CINVESTAV project
JIRA'2001/17 is acknowledged.

%%%%%%%%%%%%%%%%%%%%%%%%%%%%%%%%%%%%%

%\end{multicols}
\end{document}